\documentstyle[psfig,onecolumn]{mn2e}
\newif\ifAMStwofonts

\newcommand{\etal}{et al. }

\newcommand{\nustar}{{\it NuStar} }

\newcommand{\astroh}{{\it Astro-H} }

\newcommand{\mcarlo}{{Monte Carlo} }

\newcommand{\chandra}{{\it Chandra} }

\newcommand{\fekalfa}{{Fe~K$\alpha$} }

\newcommand{\nika}{{Ni~K$\alpha$} }
\newcommand{\nikap}{{Ni~K$\alpha$}}
\newcommand{\nikb}{{Ni~K$\beta$} }
\newcommand{\nikbp}{{Ni~K$\beta$}}

\newcommand{\nh}{$N_{\rm H}$ }
\newcommand{\nhp}{$N_{\rm H}$}

\newcommand{\thetaobs}{{$\theta_{\rm obs}$} }
\newcommand{\thetaobsp}{{$\theta_{\rm obs}$}}

\newcommand{\tablecosrange}{{Table~1} }

\newcommand{\fignikafluxvsnh}{{Fig.~1} }
\newcommand{\fignikafluxvsnhp}{{Fig.~1}}
\newcommand{\fignikaewvsnh}{{Fig.~2} }
\newcommand{\fignikaewvsnhp}{{Fig.~2}}
\newcommand{\figewgamratio}{{Fig.~3} }

\newcommand{\figcsratio}{{Fig.~4} }
\newcommand{\figcsratiop}{{Fig.~4}}
\newcommand{\fignikacshoulders}{{Fig.~5} }

\title{Monte Carlo simulations of the Nickel K$\alpha$ fluorescent 
emission line in a toroidal geometry}

\author[Tahir Yaqoob \& Kendrah D. Murphy]
{Tahir Yaqoob$^{1}$ and Kendrah D. Murphy$^{2}$ \\
$^{1}$Department of Physics and Astronomy, Johns Hopkins University, Baltimore, MD 21218. \\
$^{2}$Department of Physics, Skidmore College, 815 North Broadway, Saratoga Springs, NY 12866. \\
}

\date{Accepted 2010 November 9. Received 2010 November 9; in original form 2010 October 5}

\begin{document}

\maketitle

\begin{abstract} 
We present new results from Monte Carlo calculations of the
flux and equivalent width (EW) of the \nika fluorescent emission line
in the toroidal X-ray reprocessor model of Murphy \& Yaqoob 
(2009, MNRAS, 397, 1549).
In the Compton-thin regime, the EW of the \nika line is a factor
of $\sim 22$ less than that of the \fekalfa line but this
factor can be as low as $\sim 6$ in the Compton-thick regime. 
We show that the optically-thin limit for this ratio 
depends only on the Fe to Ni abundance ratio, it
being independent of
the geometry and covering factor of the reprocessor, and 
also independent of the shape of the incident X-ray continuum.
We give some useful analytic expressions for the 
absolute flux and the EW of
the \nika line in the optically-thin limit.
When the reprocessor is Compton-thick and the incident continuum
is a power-law with a photon index of 1.9,
the \nika EW has a maximum value of $\sim 3$~eV and $\sim 250$~eV
for non-intercepting and intercepting lines-of-sight respectively.
Larger EWs are obtained for flatter continua. 
We have also studied the Compton shoulder of the \nika line and
find that the ratio of scattered to unscattered flux in the line
has a maximum value of 0.26, less than the corresponding maximum
for the \fekalfa line. However, we find that the shape of the
Compton shoulder profile for a given column density and inclination
angle of the torus is similar to the corresponding
profile for the \fekalfa line. Our results will be useful for interpreting
X-ray spectra of active galactic nuclei (AGNs) and X-ray binary
systems in which the system parameters are favorable for
the \nika line to be detected. 

\end{abstract}


\section{Introduction}
\label{torusintro}

The \nika fluorescent emission line has the potential to offer complementary
diagnostics to the \fekalfa fluorescent line in active galactic
nuclei (AGNs) and some X-ray binary systems in which the \fekalfa
line is detected (e.g., see Torrej\'{o}n \etal 2010,
and references therein). In AGNs, 
the narrow \fekalfa emission line
is a ubiquitous feature of the X-ray spectrum of both type~1 and type~2
sources (e.g. see Shu, Yaqoob, \& Wang 2010, and references therein).
However, since the abundance of Ni is more than an order of
magnitude less than that of Fe, the \nika line is expected to
be weak. Nevertheless, it has been detected in a few AGNs.
Three of the best examples are the Circinus galaxy 
(Molendi, Bianchi, \& Matt 2003),
NGC 6552 (Reynolds \etal 1994), and NGC 1068 (Matt \etal 2004; Pounds 
\& Vaughan 2006). 
In all of these sources the equivalent width of the \fekalfa line
is large (hundreds of eV, to over 1~keV) because the X-ray
spectrum is dominated by, or has a 
relatively large contribution from, reflection in Compton-thick matter.
Naturally, such sources are the most likely to yield detections of
the \nika line because its EW will also be correspondingly larger,
in tandem with that of the \fekalfa line EW.
The \nika line has also been
detected in a source that is not reflection-dominated,
but still moderately absorbed (Centaurus~A), 
albeit with a lower statistical significance of detection
(Markowitz \etal 2007).
Improvements in the sensitivity of X-ray detectors
in the $\sim 7-10$~keV band aboard forthcoming X-ray astronomy
missions such as \nustar and \astroh will likely reveal 
detections of the \nika emission line in a larger number of 
accreting X-ray sources and will therefore open up the
opportunity to use the line as a diagnostic tool
in conjunction with the \fekalfa line. 

The results of
model calculations of the
flux and EW of a \nika fluorescent line that originates in neutral matter,
as expected in AGNs, have been reported in the
literature for disk and spherical
geometries (e.g.,  Reynolds \etal 1994; Matt, Fabian, \& Reynolds 1997).
In the present paper we study the theoretical properties
of the \nika line produced by the toroidal X-ray reprocessor model
of Murphy \& Yaqoob (2009; hereafter, MY09).
The paper is organized as follows.
In \S\ref{mytorusmodel}
we give a brief overview of the model and key assumptions.
We present the results of \mcarlo simulations for the
\nika line flux and EW in \S\ref{lineflux} and \S\ref{lineew}
respectively. In \S\ref{linecs} we show
results for the ratio of the Compton-scattered to unscattered
line flux and we discuss the Compton shoulder
of the \nika emission line.
We summarize our conclusions in \S\ref{summary}.

\section{Toroidal X-ray reprocessor model overview}
\label{mytorusmodel}

Here we give a brief overview of
our model Monte Carlo simulations and the key assumptions that
they are based upon (further details can be found in MY09).
We assume that the reprocessing material is uniform and
neutral (cold).
X-ray spectroscopy of AGNs shows overwhelming evidence for 
the narrow \fekalfa line peaking at
$\sim 6.4$~keV, indicating that the matter responsible for producing 
{\it that}
line is essentially neutral (e.g. Sulentic
\etal 1998; Weaver, Gelbord, \& Yaqoob 2001; 
Page \etal 2004; Yaqoob \& Padmanabhan 2004;
Jim\'{e}nez-Bail\'{o}n \etal 2005;  Zhou \& Wang 2005; 
Jiang, Wang, \& Wang 2006; Levenson \etal 2006;
Shu \etal 2010). Although
emission lines from ionized species of Fe are observed in some AGN
(e.g. Yaqoob \etal 2003; Bianchi \etal 2005, 2008), the present paper is
concerned specifically with modeling the \nika fluorescent emission
line that originates in the same material as the \fekalfa line 
component that is centered around 6.4~keV.
We note that this \fekalfa line at  $\sim 6.4$~keV is also
observed in some X-ray binaries (e.g., Torrej\'{o}n \etal 2010)
but it is not as common as it is in AGNs.

Our geometry is an azimuthally-symmetric
doughnut-like torus with a circular cross-section,
characterized by only two parameters, namely
the half-opening angle, $\theta_{0}$, and the
equatorial column density, \nh (see Fig.~1 in MY09).
If $a$ is the
radius of the circular cross-section of the torus, and $c+a$ is
the equatorial (i.e. maximum)
radius of the torus then $(a/c)$ is a
covering factor such that  $(a/c)=[\Delta\Omega/(4\pi)]$.
Here, $\Delta\Omega$ is the solid angle subtended by the
torus at the X-ray source, which is assumed to be
located at the center of the system, emitting isotropically.
The mean column density, integrated over all incident
angles of rays through
the torus, is then $\bar{N}_{\rm H}$~$=(\pi/4)$\nhp.
The inclination angle between the observer's line of sight and the
symmetry axis of the torus is denoted by
\thetaobsp, where \thetaobs$=0^{\circ}$ corresponds to a face-on
observing angle and \thetaobs$=90^{\circ}$
corresponds to an edge-on observing angle.
In our calculations we distribute the emergent photons in
10 angle bins between $0^{\circ}$ and $90^{\circ}$ that have
equal widths in $\cos{\theta_{\rm obs}}$, and refer
to the face-on bin as \#1, and the edge-on bin as \#10
(see \tablecosrange in MY09).

The value of $\theta_{0}$ for which we have calculated
a comprehensive set of models is $60^{\circ}$,
for $N_{\rm H}$ in the range $10^{22} \ \rm cm^{-2}$ to
$10^{25} \ \rm cm^{-2}$, valid
for input spectra with energies in the range
0.5--500~keV (see MY09 for details). 
Our model employs a
full relativistic treatment of Compton scattering,
using the full differential and total Klein-Nishina Compton-scattering
cross-sections. For
$\theta_{0}=60^{\circ}$, the solid angle subtended by the torus at the
X-ray source, $\Delta\Omega$, is $2\pi$, so that
$[\Delta\Omega/(4\pi)]=(a/c)=0.5$.

We utilized photoelectric absorption cross-sections for 30 elements as
described in Verner \& Yakovlev (1995) and
Verner \etal (1996) and
we used Anders and Grevesse (1989) elemental cosmic abundances in
our calculations.
The Thomson depth
may also be expressed in terms of the column density:
$\tau_{\rm T} = K N_{\rm
H}\sigma_{\rm T} \sim 0.809N_{24}$ where $N_{24}$ is the column density in
units of $10^{24} \rm \
cm^{-2}$.
Here, we have employed
the mean number of electrons per H atom, $\frac{1}{2}(1+\mu)$,
where $\mu$ is the mean molecular weight.
With the abundances of Anders \& Grevesse, $K=1.21656$, assuming that the number of
electrons from all other elements aside from H and He is
negligible.
The Anders \& Grevesse (1989) value for the solar Ni abundance, $A_{\rm Ni}$,
is $1.78 \times 10^{-6}$
relative to H. A more recent determination by Scott \etal (2009)
yields a value of $1.48 \times 10^{-6}$ but the statistical and
systematic uncertainties do not exclude the Anders \& Grevesse (1989) value.
We use the latter for consistency with our previous results on the \fekalfa
emission line (MY09; Yaqoob \etal 2010; Yaqoob \& Murphy 2010).

The \nika fluorescent emission line consists of two components,
$K\alpha_{1}$ and $K\alpha_{2}$, at energies $7.4782$~keV and
$7.4609$~keV respectively, and with a branching ratio of $K\alpha_{1}:K\alpha_{2}=2:1$
(Bearden 1967). These line energies are appropriate for neutral matter. 
In the Monte Carlo simulations we used a single line for
\nikap, at a rest-frame monoenergetic energy, $E_{0}$, of 7.472~keV
(obtained from weighting the $K\alpha_{1}$ and $K\alpha_{2}$ values
with the branching ratio). We used a fluorescence yield,
$\omega_{K}$, for Ni of 0.414
(see Bambynek \etal 1972) and a \nikb to \nika
ratio of 0.135
(consistent with results in  Bambynek \etal 1972).

Compared to MY09, the results in the present paper have
a substantially higher statistical accuracy
because they are based
on Monte Carlo simulations with higher numbers of
injected rays at each energy, and the calculations
employ the method of
weights (as opposed to following individual photons).
Throughout the present paper we present results for power-law
incident continua (in the range 0.5--500~keV), characterized
by a photon index, $\Gamma$, by integrating
the basic monoenergetic Monte Carlo results (Greens functions--
see MY09).

\section{\nika line flux}
\label{lineflux}

In this section we discuss the flux of \nika emission-line photons
that escape the torus without any interaction with it (the
zeroth-order, or unscattered line photons).
In \S\ref{linecs} we will discuss
the scattered component of
the line emission (the Compton shoulder).
In practice it may not actually be possible to observationally
distinguish
the zeroth-order component of an emission line from its Compton
shoulder. The finite energy resolution of the instrument and/or the
velocity broadening (of all the emission-line components)
may confuse the two blended components of a line (see
discussion in Yaqoob \& Murphy 2010 for the \fekalfa line).
The Monte Carlo results for the flux of the zeroth-order component of the
\nika emission line are shown in \fignikafluxvsnh as a
function of the equatorial column density, \nhp, for each of the 10 angle bins
in \thetaobsp. The line flux, $I_{\rm Ni \ K\alpha}$, has
been normalized
to an incident continuum that has a monochromatic flux
of 1~photon $\rm cm^{-2} \ s^{-1} \ keV^{-1}$ at 1 keV. 

\begin{figure}
\centerline{
\psfig{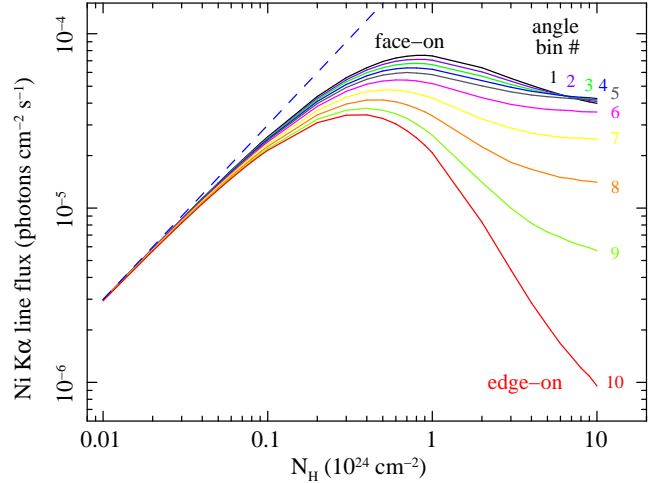}}
\caption[The \nika line flux versus \nhp.]{\footnotesize The \nika line flux
versus \nh for $\Gamma=1.9$.
Curves are shown for each of the 10 \thetaobs bins 
and are color coded and labeled by the angle bin number
(see \tablecosrange in MY09). The angle bins correspond to
equal solid angle intervals in the range $0^{\circ}$ to $90^{\circ}$.  
Angle bins 1--5 correspond to lines-of-sight that do
not intercept the torus and 
angle bins 6--10 correspond to lines-of-sight that intercept the torus.  
The normalization of the line flux corresponds to a power-law
incident continuum that has a monochromatic flux at 1 keV of
$1 \rm \ photons \ cm^{-2} \ s^{-1} \ keV^{-1}$.
The dashed line corresponds to the optically-thin limit 
for the relation between
the \nika line flux and \nh (equation~\ref{eq:thinflux}). 
}
\end{figure}

As \nh is increased, the \nika line flux first increases
but then turns over, reaching a maximum for \nh somewhere
in the range
$\sim 3-8 \times 10^{23} \ \rm cm^{-2}$, depending on the
inclination angle. This is because the escape of \nika line
photons from the medium after they are created is significantly
impeded by the absorption and scattering opacity that is
relevant at the line energy. For the edge-on angle bin the
maximum \nika line flux is attained {\it well before the
medium becomes Compton-thick}, at only $\sim 3 \times 10^{23} \ \rm cm^{-2}$.
For the face-on angle bin the maximum line flux is attained
at a higher column density ($\sim 8 \times 10^{23} \ \rm cm^{-2}$),
but still before the medium becomes Compton-thick.
The position of the turnover can be understood as approximately
corresponding to a situation when the average
optical depth to absorption plus scattering for
the zeroth-order \nika line photons is of order
unity. The behavior of the \nika zeroth-order line flux as a function of
\nh and \thetaobs is in fact very similar to that of the
flux of the \fekalfa line, for which a detailed discussion
can be found in Yaqoob \etal (2010).

\subsection{\nika line flux in the optically-thin limit}

In the optically-thin limit, 
for which absorption and scattering optical depths in the
$\sim 7-9$~keV band are $\ll 1$, we can obtain an 
approximate analytic expression
for the \nika line flux. Following Yaqoob \etal (2001),
we get

\begin{eqnarray}
I_{\rm Ni \ K\alpha}
& \sim &
8.647 \times 10^{-4} 
\ \left(\frac{\Delta\Omega}{4\pi}\right)
\ \left(\frac{\omega_{K}}{0.414}\right)
\ \left(\frac{\omega_{K\alpha}}{\omega_{K}}\right) \nonumber \\ 
& \times & \ \left(\frac{\rm A_{\rm Ni}}{1.78 \times 10^{-6}}\right) 
\ \left(\frac{\sigma_{0}}{2.86\times10^{-20} \rm \ cm^{2}}\right) \nonumber \\
& \times & \ \left(\frac{3.61}{\Gamma + \alpha -1}\right) 
\ \left(8.348 \ {\rm keV} \right)^{(1.9-\Gamma)}
\ \ \ \bar{N}_{24} \nonumber \\
& & {\rm photons \ cm^{-2} \ s^{-1}} 
\label{eq:thinflux}
\end{eqnarray}

The quantity $[\Delta\Omega/(4\pi)]$ is the fractional solid angle that 
the line-emitting matter subtends at the
X-ray source.  Note that equation~\ref{eq:thinflux} utilizes
the {\it mean (angle-averaged) column density}, not the equatorial column density.
Thus, $\bar{N}_{24} = (\pi/4)[N_{\rm H}/(10^{24} \ \rm cm^{-2})]$.
The K-shell fluorescence yield is given by
$\omega_{K}$, and $\omega_{K\alpha}$ is the yield for the
\nika line only. Using
our adopted value of 0.135 for the \nikbp/\nika line ratio,
 $\omega_{K\alpha}/\omega_{K}=0.881$.
The quantity 
$A_{\rm Ni}$ is the Ni abundance relative to Hydrogen
($1.78 \times 10^{-6}$, Anders \& Grevesse 1989).
The quantity $\sigma_{0}$
is the Ni~K shell absorption cross-section at the Ni~K 
photoelectric absorption edge threshold energy, $E_{K}$,
and $\alpha$ is the power-law index of
the cross-section as a function of energy. 
For the Verner \etal (1996) data that we have adopted,
$E_{K}=8.348$~keV, $\sigma_{0} = 2.68 \times 10^{-20} \ \rm cm^{2}$,
and $\alpha=2.71$ (obtained from fitting the K-shell cross-section
up to 30~keV with a power-law model).

The optically-thin
limit for the \nika line flux from equation~\ref{eq:thinflux}
is shown in \fignikafluxvsnh (dashed line). It can be seen that the 
Monte Carlo curves converge to this optically-thin limit, but only
for column densities $<2 \times 10^{22} \ \rm cm^{-2}$.
Note that the optically-thin limit for the \nika line flux is
{\it independent of the details of the geometry}.

We can obtain a simple result in the
optically-thin limit from equation~\ref{eq:thinflux}
for the ratio of the \fekalfa to \nika line flux. Neglecting
the small difference in the energy dependence of the
K-shell cross-section in Ni and Fe ($\sim E^{-2.71}$ and $\sim E^{-2.67}$
respectively), we have

\begin{eqnarray}
\label{eq:fenifluxratio}
\frac{I_{\rm Fe \ K\alpha}}{I_{\rm Ni \ K\alpha}} & \sim &
30.0 \left(\frac{7.124 \rm \ keV}{8.348 \ \rm keV}\right)^{(1.9-\Gamma)} 
\frac{[A_{\rm Fe}/A_{\rm Ni}]}{[A_{\rm Fe}/A_{\rm Ni}]_{\rm AG89}} \ .
\end{eqnarray}
 
In equation~\ref{eq:fenifluxratio}, 
7.124~keV is the neutral Fe K shell threshold edge energy
in Verner \etal (1996), $[A_{\rm Fe}/A_{\rm Ni}]_{\rm AG89}$ is
the Anders \& Grevesse (1989) Fe to Ni abundance ratio (26.3), 
and $[A_{\rm Fe}/A_{\rm Ni}]$ 
is the actual Fe to Ni abundance ratio in the source. What is
interesting about equation~\ref{eq:fenifluxratio}
is that not only is it independent of geometry, it is {\it independent of
the covering factor}.

\section{The \nika line equivalent width}
\label{lineew}

In \fignikaewvsnhp, we show the EWs of the
unscattered (zeroth-order) component of the
\nika line as a function of the column
density of the torus, \nhp, calculated for $\Gamma=1.9$.
The lower set of curves show the results for the non-intercepting angle
bins, and the upper set of curves show the results for the
intercepting angle bins, as indicated by the color-coded angle
bin numbers.
It can be seen in \fignikaewvsnh that
inclination-angle effects become important 
for \nh greater 
than $\sim 5\times10^{22} \ \rm cm^{-2}$.
For inclination angles that do not intercept the torus, the EW peaks between $\sim8\times10^{23} \rm
\ cm^{-2} \ and \ 10^{24}\ cm^{-2}$, then decreases by more 
than $50$\% of this peak value at \nh$=
10^{25}\ \rm cm^{-2}$. For these non-intercepting lines-of-sight,
the peak value for the EW of the \nika line, for $\Gamma=1.9$, is $\sim3$ eV.
For the lines-of-sight that intercept the torus, the \nika line EW
reaches its maximum value between 
$N_{\rm H} \sim 2-4 \times 10^{24} \ \rm cm^{-2}$,
becoming as high as $\sim 250$~eV for the edge-on angle bin.
Overall, the behavior of the EW as a function of \nh and \thetaobs is
analogous to that of the \fekalfa line, for which a detailed 
discussion can be found in MY09.

\begin{figure}
\centerline{
\psfig{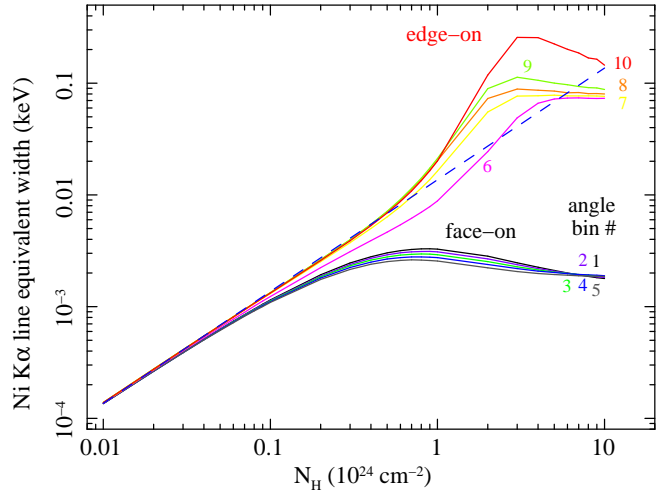}}
\caption[The \nika line equivalent width versus \nhp.]{\footnotesize
The \nika line equivalent width (EW) versus \nh for $\Gamma=1.9$.
Curves are shown for each of the 10 \thetaobs bins 
and are color coded with the same scheme as
in \fignikafluxvsnhp, and labeled by the angle bin number
(see \tablecosrange in MY09). The angle bins correspond to
equal solid angle intervals in the range $0^{\circ}$ to $90^{\circ}$.  
Angle bins 1--5 correspond to lines-of-sight that do
not intercept the torus and 
angle bins 6--10 correspond to lines-of-sight that intercept the torus.  
The dashed line corresponds to the optically-thin limit for the relation
between the \nika line EW and \nh (equation~\ref{eq:thinew}).
}
\end{figure}

\subsection{\nika line EW in the optically-thin limit}

Overlaid on the curves in \fignikaewvsnh is the theoretical 
optically-thin limit (dashed line), given by dividing
equation~\ref{eq:thinflux} by $E^{-\Gamma}$ 
(recall that the line flux in equation~\ref{eq:thinflux} is
normalized to a power-law continuum normalization at 1~keV of
1 $\rm photon \ cm^{-2} \ s^{-1} \ keV^{-1}$). Thus,
the EW of the \nika line in the optically-thin limit is 

\begin{eqnarray}
\label{eq:thinew}
EW_{\rm Ni \ K\alpha} & \sim & 39.48
\ \left(\frac{\Delta\Omega}{4\pi}\right)
\ \left(\frac{\omega_{K}}{0.414}\right)
\ \left(\frac{\omega_{K\alpha}}{\omega_{K}}\right) \nonumber \\ 
& \times & \left(\frac{\rm A_{\rm Fe}}{1.78 \times 10^{-6}}\right) 
\ \left(\frac{\sigma_{0}}{2.86\times10^{-20} \rm \ cm^{2}}\right) \nonumber \\
& \times &  \left(\frac{3.61}{\Gamma + \alpha -1}\right) 
\ \left(\frac{E_{0}}{E_{K}}\right)^{(\Gamma-1.9)} 
\ \bar{N}_{24} \ \ \ \rm eV \ .
\end{eqnarray}

As in equation~\ref{eq:thinflux}, the column density in equation~\ref{eq:thinew}
is the {\it mean, angle-averaged} column density
($\bar{N}_{24} = (\pi/4)[N_{\rm H}/(10^{24} \ \rm cm^{-2})]$).
In equation~\ref{eq:thinew}
$E_{0}$ is the \nika line centroid energy. The ratio $(E_{0}/E_{K})$
is $(7.472/8.348)=0.895065$. This happens to be very similar to
the corresponding ratio for the \fekalfa line, $(6.400/7.124)=0.898372$.
This leads to a very simple expression for the approximate ratio between
the EW of the \fekalfa line and the EW of the \nika line in the 
optically-thin limit, that
is {\it independent of the shape of the intrinsic continuum}.
In analogy to equation~\ref{eq:fenifluxratio}, we get

\begin{eqnarray}
\label{eq:feniewratio}
\frac{EW_{\rm Fe \ K\alpha}}{EW_{\rm Ni \ K\alpha}} & \sim &
22.2 \frac{[A_{\rm Ni}/A_{\rm Fe}]}{[A_{\rm Ni}/A_{\rm Fe}]_{\rm AG89}} \ .
\end{eqnarray}

As was the case for the line {\it flux} ratio, the EW ratio in the optically-thin
limit is independent of geometry {\it and} the covering factor.
Note that in the Compton-thick regime, for lines-of-sight that
intercept the torus, the \nika line EW is larger relative
to the \fekalfa line EW than simple linear 
scaling of the optically-thin case. In other words,
the EW ratio in equation~\ref{eq:feniewratio} becomes smaller as \nh increases, for
non-intercepting inclination angles. We find that the ratio has its
smallest value ($\sim 6$) for $N_{\rm H} \sim 2-3 \times 10^{24} \ \rm cm^{-2}$
and an edge-on inclination angle.

We note another important aspect of the \nika line EW versus
\nh curves in \fignikaewvsnhp. That is, fortuitously, the 
relation for the optically-thin
limit (dashed line) happens to give excellent agreement for
the edge-on inclination angle bin all the way up to 
$N_{\rm H}=6 \times 10^{23} \ \rm cm^{-2}$. This gives a very
convenient way to analytically estimate the EW of the \nika line
for an edge-on orientation and a column density less than
$6 \times 10^{23} \ \rm cm^{-2}$.  

We find that smaller values of $\Gamma$ yield larger values of the \nika EW; this is expected as there
are relatively more photons in the continuum above the Ni~K edge for flatter spectra.  \figewgamratio
shows the ratio of the \nika line EW for $\Gamma=1.5$ to the corresponding EW 
for $\Gamma=2.5$, versus \nhp, for each of the 10
inclination-angle bins (see \tablecosrange in MY09).  
In the optically-thin regime, this ratio can simply be obtained by
evaluating equation~\ref{eq:thinew} for each value of $\Gamma$ and taking the
ratio. We get a value of 1.465, and this is shown in \figewgamratio (dashed line),
from which it can be seen that there is excellent agreement with the \mcarlo
results. For the non-intercepting angle bins this ratio does not increase above $\sim 1.72$
even in the Compton-thick regime. However, for the edge-on angle bin, the ratio has its
maximum value (with respect to all the angle bins and \nh values) of $\sim 2.5$,
for $N_{\rm H} \sim 8 \times 10^{24} \ \rm cm^{-2}$.

\begin{figure}
\centerline{
\psfig{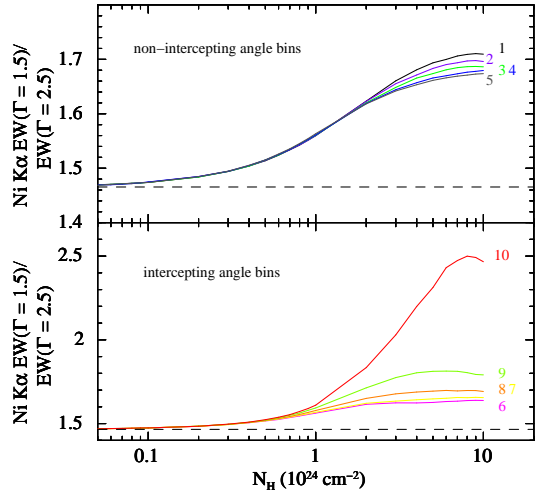}}
\caption[EW ratio for different Gammas]{\footnotesize Ratios of the \nika line equivalent width (EW) for $\Gamma=1.5$ to the corresponding
EW for $\Gamma=2.5$, versus \nhp. 
Curves are shown for each of the 10 \thetaobs bins 
and are color coded with the same scheme as
in \fignikafluxvsnh and \fignikaewvsnhp, and labeled by the angle bin number
(see \tablecosrange in MY09). The angle bins correspond to
equal solid angle intervals in the range $0^{\circ}$ to $90^{\circ}$.  
The upper panel shows angle
bins 1--5, corresponding to lines-of-sight that do
not intercept the torus, and the lower panel shows angle 
bins 6--10 corresponding to lines-of-sight that intercept the torus.  
Note that the vertical axis scale is different for the two panels.
The dashed line
shows the optically-thin limiting value of the EW ratio, obtained
from equation~\ref{eq:thinew}. }
\end{figure}

The EW versus \nh curves have an explicit dependence on the assumed opening angle of the torus.  This is
because different opening angles correspond to different solid angles subtended by the torus at the source
and to different projection-angle effects.  In the optically-thin regime, this dependence is linear.
In the Compton-thick regime, there is a more complicated dependence that must be determined
by additional Monte Carlo simulations, which will be the subject of future investigation.

\section{\nika line Compton shoulder}
\label{linecs}

In addition to the zeroth-order (unscattered) core of the \nika emission line, 
the shape and relative magnitude of
the scattered component of the \nika emission
line (i.e. the Compton shoulder)
are also sensitive to the properties of the reprocessor
(e.g., see Sunyaev \& Churazov 1996;
Matt 2002; Watanabe \etal 2003; Yaqoob \& Murphy 2010).  
\figcsratio shows plots of the ratio of the
total number of scattered \nika 
line photons to zeroth-order
\nika line photons (hereafter, CS ratio) versus \nhp.  
The ratios are shown for an input power-law continuum with
$\Gamma=1.9$, for the face-on and edge-on inclination angle bins
(see \tablecosrange in MY09).
It can seen that the CS ratio peaks at $N_{\rm H} \sim 2-3
\ \times 10^{24} \ \rm cm^{-2}$,
reaching a maximum of $\sim 0.22$ (face-on),
and $\sim 0.26$ (edge-on).
These are $\sim 75\%$ and $\sim 70\%$ of the corresponding
ratios for the \fekalfa line (see MY09 and Yaqoob \& Murphy 2010).
For the face-on inclination angle, the CS ratio for the \nika line remains at $\sim 0.22$
once the maximum is reached (even if the column density is increased further)  since
the Compton shoulder photons escape from
within a Compton-depth or so from the illuminated surfaces
of the torus for lines-of-sight that are not obscured.
For the edge-on inclination angle the CS ratio for the \nika line declines as a function
of column density after reaching its maximum value, due to a higher
probability of absorption at higher column densities.

\begin{figure}
\centerline{
\psfig{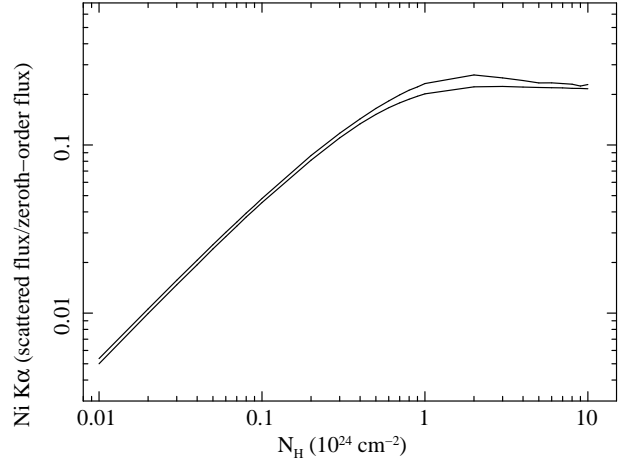}}
\caption[CS ratio for Ni Ka line]{\footnotesize The ratios of the total number of scattered \nika line photons to
the number of zeroth-order \nika
line photons, versus \nhp.  Ratios are shown for an input power-law continuum with $\Gamma=1.9$, for
 the face-on inclination angle bin (lower curve) and the 
edge-on inclination angle bin (upper curve). }
\end{figure}

We found that for all values of \thetaobs for the torus,
there was no detectable difference in the CS ratio as a function
of $\Gamma$ up to the \nh value that gives the maximum CS ratio
(for a given value of \thetaobsp). After that, the CS ratios
diverge for different values of $\Gamma$, with flatter incident
continua giving larger CS ratios. For the face-on inclination angle
the CS ratio at $N_{\rm H} = 10^{25} \ \rm cm^{-2}$ varies
between $\sim 0.21$ to $\sim 0.22$ as $\Gamma$ varies from
2.5 to 1.5. For the edge-on case, the CS ratio at $N_{\rm H} = 10^{25} \ \rm cm^{-2}$ varies
between $\sim 0.22$ to $\sim 0.23$ as $\Gamma$ varies from
2.5 to 1.5. Flatter spectra have relatively more continuum photons
at higher energies so that the \nika line photons
are produced deeper in the medium, increasing the average
Compton depth for zeroth-order line photons to scatter before escaping.
These variations in the CS ratio with $\Gamma$ are likely to
be too small to be detectable in practice.

The {\it shape} of the Compton shoulder of a fluorescent
emission-line escaping from the torus also has a dependence on
the column density and inclination angle of the torus.
We found that the shapes of the Compton shoulder profiles
for the \nika line are practically indistinguishable
from the shapes of the \fekalfa line Compton shoulder profiles
(see MY09 and Yaqoob \& Murphy 2010).
\fignikacshoulders
illustrates the shapes of the \nika line Compton shoulder
(solid lines) for a power-law
incident continuum with $\Gamma=1.9$, for two column densities
($10^{24}$ and $10^{25} \rm \ cm^{-2}$) and 
two inclination angles of the torus (face-on and edge-on).
Corresponding Compton shoulder profiles are also shown
for the \fekalfa line (dotted lines) for comparison.
The Compton shoulder shapes shown in \fignikacshoulders 
have no velocity broadening applied to
them. The Compton shoulders are shown
in wavelength space in units of the dimensionless Compton
wavelength shift with respect to the zeroth-order rest-frame
energy of the emission line. In other words, if $E$ is the energy
of a line photon, and $E_{0}$ is the zeroth-order line energy,
$\Delta \lambda = (511 \ {\rm keV}/E)-(511 \ {\rm keV}/E_{0})$.
In order to facilitate a direct comparison of the Compton
shoulder profile shapes for different column densities and
inclination angles, all of the profiles in \fignikacshoulders have
been normalized to a total flux of unity. It should be remembered that
the absolute flux of the Compton shoulder varies
significantly with column density, and the flux ratio for
two column densities can be estimated using \figcsratiop.

\begin{figure}
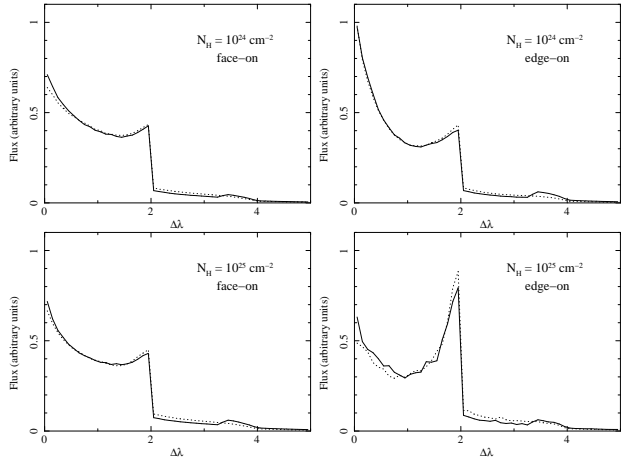

\centerline{
\psfig{figure=f5a.ps,height=3.0cm,angle=270}
\psfig{figure=f5b.ps,height=3.0cm,angle=270}
}
\centerline{
\psfig{figure=f5c.ps,height=3.0cm,angle=270}
\psfig{figure=f5d.ps,height=3.0cm,angle=270}
}
\caption[Shapes of the Compton shoulder for different column densities and inclination angles.] {\footnotesize
The \nika emission-line Compton shoulders (solid curves) for a power-law
incident continuum with $\Gamma=1.9$, for two column densities
and two inclination angle bins, as indicated
($N_{\rm H} = 10^{24}$ and $10^{25} \rm \ cm^{-2}$,
each for face-on and edge-on orientations of the torus).
The dotted curves show the corresponding Compton shoulder
profiles for the \fekalfa emission line, for the same
column densities and inclination angles.
No velocity broadening has been applied.
{\it Note that in order to directly compare the Compton shoulder
shapes, the total flux for each shoulder has been renormalized to the same
value}. The line flux (in units of normalized flux per unit
wavelength shift) is plotted
against the dimensionless Compton
wavelength shift with respect to the zeroth-order rest-frame
energy of the emission line ($E_{0}$), $\Delta \lambda =
(511 \ {\rm keV}/E)-(511 \ {\rm keV}/E_{0})$.
}
\end{figure}

Yaqoob \& Murphy (2010) discussed the dependence of the shape of the \fekalfa line
Compton shoulder on \nh and inclination angle in considerable detail.
Differences in the shapes of the Compton shoulder profiles for the \nika line
and the \fekalfa line for the same model parameters only become
apparent for $N_{\rm H} \gg 10^{24} \ \rm cm^{-2}$ and edge-on inclination
angles. However, even at
$N_{\rm H} =  10^{25} \ \rm cm^{-2}$, the differences are less than $15\%$.
Therefore, since the shapes of the Compton shoulder profiles for the \nika line
are similar to the \fekalfa line Compton shoulder profiles
within the statistical uncertainties of the Monte Carlo simulations,
we do not discuss the \nika line Compton shoulder further.
The discussion and interpretation of the \fekalfa line Compton shoulder
in Yaqoob \& Murphy (2010) can be applied to the \nika line.

\section{Summary}
\label{summary}

We have presented some new results for the flux and EW of the
\nika fluorescent emission line from \mcarlo simulations of
a toroidal reprocessor illuminated by a power-law X-ray continuum.
Our results cover values of the equatorial column density, \nhp, of
$10^{22} \rm \ cm^{-2}$ to $10^{25} \rm \ cm^{-2}$, and the calculations
were performed for a global covering factor of 0.5 and 
cosmic elemental abundances. As might be expected, the
behavior of the \nika line
flux and EW as a function of the column density and inclination
angle of the torus is similar to that of the \fekalfa line.
However, the EW of the \nika line is a factor of $\sim 22$
smaller than that of the \fekalfa in the Compton-thin regime.
In the Compton-thick regime, the EW of the \nika line
reaches a maximum of $\sim 3$~eV for lines-of-sight that do not
intercept the torus. For intercepting lines-of-sight the \nika
EW can be as high as $\sim 250$~eV.
The ratio of the \fekalfa to \nika line EW in the Compton-thick
regime, for intercepting lines-of-sight, can be significantly
less than the optically-thin limit, as low as $\sim 6$.
The above results pertain to an incident power-law X-ray
continuum with a photon index of 1.9. Flatter continua
give larger EWs and steeper continua give smaller EWs.
Varying $\Gamma$ in the range $\Gamma=1.5$ to 2.5 can
change the \nika EW by up to $\sim 70\%$ in the Compton-thick 
regime.

We have given analytic expressions for the \nika flux and
EW in the optically-thin limit. We have also given simple
analytic expressions, in the optically-thin limit,
for the ratio of the \fekalfa to \nika line flux, as well
as the ratio of the \fekalfa to \nika line EW. Both of these
ratios are independent of the geometry and covering factor
of the reprocessor. Moreover, we have found that
the ratio of the \fekalfa to \nika line EW is independent of
$\Gamma$, depending only on the Fe to Ni abundance ratio
(in the optically-thin limit). 

We have also investigated the Compton shoulder of the
\nika line and we found that the ratio of the flux  in the
Compton shoulder to that in the zeroth-order component of the
line has a maximum value of $\sim 0.22$ and
$\sim 0.26$ for face-on and edge-on inclination angles
respectively. These are less than the corresponding
maxima for the \fekalfa line. However,
we have found that the {\it shapes} of the \nika and
\fekalfa line Compton shoulder profiles are indistinguishable
within the statistical accuracy of the \mcarlo results,
except for edge-on inclination angles and $N_{\rm H} \gg 10^{24}
\ \rm cm^{-2}$. However, even for \nh as high as $10^{25}
\ \rm cm^{-2}$, the differences are less than $15\%$.

Our calculations of the 
\nika line flux, EW, and Compton shoulder are meant to
serve as a baseline reference because the detailed results, especially
in the Compton-thick regime, depend on a number of factors
that have not been investigated here. The opening angle
of the torus (or effective covering factor) may of
course be different to the value used in the calculations. 
Also, as suggested by Lubinski \etal (2010), part of the
torus may be shielded from the
X-ray continuum by the accretion disk. However, quantifying this
is subject to uncertainties in the geometry of the X-ray source and
accretion disk system. Further deviations from the baseline
model could occur if the torus does not
have a circular cross-section and/or if the torus is clumpy
(e.g., Krolik \& Begelman 1988; Nenkova, Ivezi\'{c} \& Elitzur 2002).
Extension of the parameter space for our model will be the subject
of future work.

Acknowledgments \\
Partial support (TY) for this work was provided by NASA through \chandra Award
TM0-11009X, issued by the Chandra X-ray Observatory Center,
which is operated by the Smithsonian Astrophysical Observatory for and
on behalf of the NASA under contract NAS8-39073.
Partial support (TY) from NASA grants NNX09AD01G and NNX10AE83G is also
acknowledged. The authors thank Andrzej Zdziarski for helpful
comments for improving the paper.

\end{document}